\documentclass[aps,pra,groupedaddress]{revtex4}
\usepackage{graphicx,subfigure}
\usepackage{epsfig}

\usepackage{bbold}

\begin{document}
\title{Entanglement in two site Bose-Hubbard model}
\author{K. V. S. Shiv Chaitanya}
\email[]{E-mail: chaitanya@hyderabad.bits-pilani.ac.in}
\affiliation{BITS Pilani, Hyderabad Campus, Jawahar Nagar, \\Shameerpet Mandal,
Hyderabad, India 500 078.}
\author{ Sibasish Ghosh }
\email[]{E-mail: sibasish@imsc.res.in}
\affiliation{Optics $\&$ Quantum Information Group,\\The Institute of Mathematical Sciences, C.I.T Campus\\Taramani, Chennai, India, 600113.}
  \author{V. Srinivasan}
  \email[]{ E-mail:: vsspster@gmail.com}
\affiliation{Department of Theoretical Physics, University of Madras, \\ Guindy, Chennai, India, 600025.}

\begin{abstract}
In this paper, we study the decoherence and entanglement properties for the two site Bose-Hubbard model
in the presence of a non-linear damping. We apply the techniques of thermo field dynamics and then use 
Hartree-Fock approximation to solve the corresponding master equation. The expectation values of the approximated field operators appearing in the solution of master equation, are 
computed self-consistently.  We solve this master equation for a small time $t$ so that we get 
the analytical solution, thereby we compute the decoherence and entanglement properties of the solution of the
two-mode bosonic system.
\end{abstract}

\maketitle
\section{Introduction}
In recent years, there is lot of interest generated in the study of entanglement properties of ultra cold atoms
\cite{ula1,ula2,ula3,ula4,ula5,ula6}. In one such study, the  single-site addressability in a two-dimensional optical lattice \cite{qie} has been demonstrated which could be a natural resource for applications of quantum information processing with neutral atoms. 
 In all the 
experimental demonstrations of ultra cold atoms, loss is an important role which gives rise to decoherence and in turn destroying the quantum correlations. The losses due to decoherence can be modelled by a master equation.  One such model is examined in the ref \cite{ula5, ula6} for a linear damping  using the Bose-Hubbard model.
The Bose-Hubbard model \cite{ci} is one of the popular model used to study the evolution of cold atoms and the Bose-Einstein condensates in an optical lattice. In this paper, we examin the two-site Bose-Hubbard model to study the entanglement and decoherence properties of two mode states under the action of non-linear damping.  We consider the following master equation for density matrix $\rho$
in a non-linear medium
\begin{eqnarray}
\frac{\partial}{\partial t}\rho=\frac{i}{\hbar}[H,\rho]+\kappa\sum_{k=1}^{K}([a_kb_k,\rho a_k^\dagger b_k^\dagger]
+[a_kb_k\rho, a_k^\dagger b_k^\dagger])\label{mbh}
\end{eqnarray}
here $\kappa$ is a damping coefficient, $a_k$ and $b_k$ bosonic annihilation operators referring to
atoms in the internal states $\vert N_1\rangle$ and $\vert N_2\rangle$, respectively, with one boson in the kth lattice site and $K$ is the number of lattice sites and $H$ is the Hamiltonian for the Bose-Hubbard model which describes the optical lattice. In this paper, we are studying the model in the presence of non-linear damping corresponding to the term associated with $\kappa$.

For solving this master equation we use the techniques of thermo field dynamics 
and thereby the Hartree-Fock approximation method to convert the two-site Bose-Hubbard model in eq (\ref{mbh}) into a two-mode bosonic system.
The two-site Bose-Hubbard model is used to study
Josephson tunnelling between two Bose-Einstein condensates (BEC). This toy model can be used to study the BEC.
The expectation values of the approximated field is computed self-consistently.  
We solve the master equation (\ref{mbh}) for a small time $t$ so that we get 
the analytical solution, thereby we compute the decoherence and entanglement properites of the
two-mode bosonic system.

The thermo field dynamics (TFD)\cite{U1,U2,U3,U4,U5} is a finite temperature field theory. It is 
applied to many branches in high energy physics \cite{U2} and many-body systems \cite{U5}.
The thermo field dynamics  is used
to solve the master equation by in presence of Kerr medium \cite{vs1,vs2} using disentanglement
theorem for any arbitrary initial conditions.  This
formalism, presented in ref \cite{vs1,vs2,kap,vs3,vs4} has two sailent features, first,  solving the master equation is reduced
to solving a Schr\"odinger equation, thus all the techniques available to solve the
Schr\"odinger equation are applicable here. Second, the thermal coherent state under
the master equation evolution goes over to a thermal coherent state.

A brief description of TFD is given in appendix D. In TFD, any master equation is given by
\begin{eqnarray}
\frac{\partial}{\partial t}\vert \rho(t)\rangle = -i\hat{H}\vert \rho\rangle\label{sc}
\end{eqnarray}
where $\vert \rho\rangle$ is a vector in an extended Hilbert space ${\cal H}\otimes {\cal H}^*$ and
\begin{eqnarray}
 -i\hat{H}=i(H-\tilde{H})+L.
\end{eqnarray}
where $H$ is the Hamiltonian in Hilbert space ${\cal H}$,  $\tilde{H}$ is the  Hamiltonian in Hilbert space ${\cal H}^*$ and L is the Liouville term, as mentioned in appendix D .
Thus, $-i\hat{H}$ is tildian and the problem of solving master equation is reduced to solving a Schr\"odinger like equation namely eq (\ref{sc}). Then, the symmetry associated to the Hamiltonian such as $su(1,1)$ symmetry, are exploited to solve the eq (\ref{sc}) and thereby the master equation.

In this paper, we follow this formalism  to solve the master equation(\ref{mbh}) for the two site Bose-Hubbard model in the presence of  non-linear medium. 

This paper is organized as follows. In the section II, we describe the two site Hubbard model. In section III, we describe the solution of corresponding master equation. In section III-A, we discuss the self consistency analysis of the solution for a short initial time. The issue of entanglement of the solution - two mode Gaussian state - is described in section IV. While the issue of decoherence is discussed in section V. In the section VI, we draw conclusions. A few appendices are added to provide the details.

\section{Two site Bose-Hubbard Model}
The Hamiltonian $H$ of the Bose-Hubbard model, appeared on the RHS of eq (\ref{mbh}), describes the interaction of bosons situated on an optical lattice, and is given by 
\begin{eqnarray}
H &=&\omega\sum_{k}(a_k^\dagger a_k+b_k^\dagger b_k)
-J\sum_{\{k,l\}}(a_k^\dagger b_l+b_k^\dagger a_l)
\nonumber\\&&+\frac{U_a}{2}\sum_{k}a_k^{\dagger}a_k^{\dagger}a_ka_k+\frac{U_b}{2}\sum_{k}b_k^{\dagger}b_k^{\dagger}b_kb_k \nonumber\\&&+ \frac{U_{ab}}{2}\sum_{\{k,l\}}a_k^{\dagger}a_kb_l^{\dagger}b_l
\label{BH}
\end{eqnarray}
with $a_k$ and $b_k$ bosonic annihilation operators referring to
atoms in the internal states $\vert N_1\rangle$ and $\vert N_2\rangle$, respectively, with one boson in the $k$th lattice site, $K$ is the number of lattice sites and $\{k,l\}$ refers to the adjacent lattice points $k$ and $l$.
The interaction term $J$ in the Hamiltonian describes the induced hopping between adjacent cells,  $\omega$
is the frequency of the atom in the lattice. The on-site interactions of atoms are described by the interaction strengths $U_a$
and $U_b$, and a nearest-neighbour interaction by $U_{ab}$. For further details see ref \cite{ci}. 
To study the decoherence and the entanglement properties of Bose-Hubbard model, for simplicity,
we consider the toy model, in which the Bose-Hubbard model is written for the two site
interaction only (with the respective bosonic annihilation operators being $a$ and $b$). The master equation (\ref{mbh}) for the two site Bose-Hubbard Hamiltonian $H$ (given in (\ref{BH})) is given by 
\begin{eqnarray}
\frac{\partial}{\partial t}\rho &=&
-i\omega(a^\dagger a\rho-\rho a^\dagger a)
-iJ(a^\dagger b\rho-\rho a^\dagger b)\nonumber\\&&-iJ(b^\dagger a\rho-\rho b^\dagger a)
+i\frac{U_a}{2}(a^{\dagger}a^{\dagger}aa\rho-\rho a^{\dagger}a^{\dagger}aa)\nonumber\\&&+i\frac{U_b}{2}(b^{\dagger}b^{\dagger}bb\rho-\rho b^{\dagger}b^{\dagger}bb)-i\omega(b^\dagger b\rho-\rho bb^\dagger)
\nonumber\\&&+i\frac{U_{ab}}{2}(a^{\dagger}b^{\dagger}ab\rho-\rho a^{\dagger}b^{\dagger}ab)\nonumber\\&&+\frac{\kappa}{2}
\left(2ab\rho a^{\dagger}b^{\dagger}
- a^{\dagger}b^{\dagger}ab\rho-\rho a^{\dagger}b^{\dagger}ab\right).
\label{BHu}
\end{eqnarray}
At first we consider the special case to solve this master equation in which $J=U_a=U_b=0$ and $U_{ab}=U$ (say), which corresponds to the Mott insulating phase. Then the  
master equation (\ref{BHu}) reduces to 
\begin{eqnarray}
\frac{\partial}{\partial t}\rho &=&
-i\omega(a^\dagger a\rho-\rho a^\dagger a)
-i\omega(b^\dagger b\rho-\rho bb^\dagger)
\nonumber\\&&+i\frac{U}{2}(a^{\dagger}b^{\dagger}ab\rho-\rho a^{\dagger}b^{\dagger}ab)\nonumber\\&&+\frac{\kappa}{2}
\left(2ab\rho a^{\dagger}b^{\dagger}
- a^{\dagger}b^{\dagger}ab\rho-\rho a^{\dagger}b^{\dagger}ab\right).
\label{pair}
\end{eqnarray}
Now, we apply the thermo field dynamics techniques to convert the master equation (\ref{pair}) into a Schr\"odinger equation
by applying $\vert I\rangle$ from the right to the eq (\ref{pair}):
\begin{equation}
\frac{\partial}{\partial t}\vert\rho\rangle=-i\hat{H} \vert\rho\rangle,\label{hal}
\end{equation}
where $\vert\rho\rangle$ pure state in the Hilbert space ${\cal H}\otimes {\cal H}^*$ and the Hamiltonian $\hat{H}$ is given by
\begin{eqnarray}
-i\hat{H}  &=&
-i\omega(a^{\dagger}a-\tilde{a}\tilde{a}^{\dagger})
-i\omega(b^{\dagger}b-\tilde{b}\tilde{b}^{\dagger})
\nonumber\\&&+i\frac{U}{2}(a^{\dagger}b^{\dagger}ab
-\tilde{a}\tilde{b}\tilde{a}^{\dagger}\tilde{b}^{\dagger})\nonumber\\&&
+\frac{\kappa}{2}
\left(2ab\tilde{a}\tilde{b}
- a^{\dagger}b^{\dagger}ab-\tilde{a}\tilde{b}\tilde{a}^{\dagger}\tilde{b}^{\dagger}\right),
\label{pair1}
\end{eqnarray}
here the $a$, $b$ are the annihilation operators act on the ${\cal H}$ and  $ \tilde{a}$ and $ \tilde{b}$ are the annihilation operators act on the Hilbert space ${\cal H}^*$ ( for detail see appendix D).

This master equation (\ref{pair1}) is a non-linear equation (because $(U,\kappa)\neq(0,0)$ here), and in general it is difficult to get an analytical
solution of this equation. A way out would be to apply the Hartree-Fock approximation and treat the approximated field
self-consistently. In the case of thermo field dynamics, a selfconsistent theory using Hartree-Fock approximation is developed for non linear master equations in presence of the non-linear medium in ref \cite{vs4}.
Applying, Hartree-Fock approximation for each term in (\ref{pair1}) as follows
\begin{eqnarray}
 2\tilde{a}^{\dagger}\tilde{b}^{\dagger}\tilde{a}\tilde{b}
 =\tilde{a}^{\dagger}\tilde{b}^{\dagger}\langle \tilde{a}\tilde{b}\rangle +
 \langle \tilde{a}^{\dagger}\tilde{b}^{\dagger}\rangle \tilde{a}\tilde{b},\;
2a^{\dagger}b^{\dagger}ab=\langle a^{\dagger}b^{\dagger}\rangle ab+
a^{\dagger}b^{\dagger}\langle ab\rangle ,\;
2\tilde{a}\tilde{b}ab=\tilde{a}\tilde{b}\langle ab\rangle + ab\langle \tilde{a}\tilde{b}\rangle,
\end{eqnarray}
with  $\langle ab\rangle=\langle \tilde{a}\tilde{b}\rangle=\Delta(t)$, the Hamiltonian in (\ref{pair1}) is decoupled into tildien and non tildian parts
\begin{eqnarray}
\hat{H}  &=&(H_1+H_2),\label{ji}
\end{eqnarray}
where
\begin{eqnarray}
H_1  &=&\omega (a^{\dagger}a+ b^{\dagger}b) +\frac{i\kappa\Delta(t)}{4}
\left(ab - a^{\dagger}b^{\dagger}\right)\nonumber\\&&-\frac{U\Delta(t)}{4}(a^{\dagger}b^{\dagger}+ab)\label{sq}\end{eqnarray}
and
\begin{eqnarray}H_2&=&
-\omega(\tilde{a}\tilde{a}^{\dagger}
+\tilde{b}\tilde{b}^{\dagger})
+\frac{i\kappa\Delta(t)}{4}
\left(\tilde{a}\tilde{b}- \tilde{a}^{\dagger}\tilde{b}^{\dagger}\right)
\nonumber\\&&-\frac{U\Delta_1(t)}{4}(\tilde{a}^{\dagger}\tilde{b}^{\dagger}
+\tilde{a}\tilde{b}).
\label{sq1}
\end{eqnarray}
 The solution of (\ref{hal}) is then given by
\begin{equation}
\vert \rho (t)\rangle =\big( exp[-i\int dt H_1]\otimes exp[-i\int dt H_2]\big)\vert \rho (0)\rangle,\label{soll}
\end{equation}
where $\vert\rho(0)\rangle$ is an initial state in ${\cal H}\otimes {\cal H}^*$. 

It is clear from the above that the two Hamiltonians $H_1$ and $H_2$ are independent in the sense that $H_1$ is acting on non-tildian system and $H_2$ is acting on tildian system
Hence, we can work with one of the Hamiltonians and similar thing works for the other Hamiltonian 
(exept for interchanging between $\omega$ and $-\omega$).

To study the decoherence and entanglement properties of the two-mode states under the action of the master equation (\ref{pair}),(equivalently, the Schr\"odinger equation (\ref{hal})), we would like to exploit the underlying 
symmetry associated with the Hamiltonians (\ref{sq}) and (\ref{sq1}).
This is accomplished by defining the following operators :
\begin{eqnarray}
\mathcal{N}&=&a^\dagger a + b^\dagger b, \; \mathcal{K}_+=a^\dagger b^\dagger, \;\mathcal{K}_-=a b\label{k1}\\
\tilde{\mathcal{N}}&=&\tilde{a}\tilde{a}^{\dagger}
+\tilde{b}\tilde{b}^{\dagger}, \; \tilde{\mathcal{K}}_+=\tilde{a}^{\dagger}\tilde{b}^{\dagger}, \;\tilde{\mathcal{K}}_-=\tilde{a}\tilde{b}\label{k2}
\end{eqnarray}
which satisfy the $su(1,1)$ algebra
\begin{eqnarray}
[\mathcal{N},\mathcal{K}_+ ]&=&\mathcal{K}_+, \; [\mathcal{N},\mathcal{K}_- ]=\mathcal{K}_-, \;
[\mathcal{K}_+,\mathcal{K}_- ]=2\mathcal{N}.\label{comu}
\end{eqnarray}
Similar algebra holds for tildians operators. Rewriting the Hamiltonian (\ref{sq}) in terms of the $su(1,1)$ generators one gets
\begin{eqnarray}
-iH_1  &=&-i\omega \mathcal{N}+\frac{\kappa\Delta(t)}{2}
\left(\mathcal{K}_- - \mathcal{K}_+\right)+i\frac{U\Delta(t)}{2}(\mathcal{K}_+ + \mathcal{K}_-),
\label{sq11}
\end{eqnarray}
and similarly the Hamiltonian (\ref{sq1}). It is clear that these two Hamiltonians are associated with the $su(1,1)$ symmetry. Hence, the underlying symmetry of the Schr\"odinger equation (\ref{hal}) is $su(1,1)\times su(1,1)$. One of the important features of this $su(1,1)$ symmetry is that it gives rise to squeezing, and in turn, it gives rise to entanglement. If the initial state $\rho(0)$ is a two-mode Gaussian state then, under the action of two-mode squeezing, the final state turns out to be also a Gaussian state. Thus, one can use the separability  criterion for Gaussian states \cite{sim} to check for the entanglement of the time evolved state $\rho(t)$.

\section{Solution of Master Equation}
To get the solution to the Schr\"odinger equation (\ref{hal}) we use the Hartree-Fock 
approximation and treat the approximated field as background field, and the later has to be computed 
self-consistently in order to get the solution of (\ref{hal}). For doing a self-consistent analysis we first exactly diagonalize the
Hamiltonians in eqs (\ref{sq}) and (\ref{sq1}) with the help of the underlying $su(1,1)$ symmetry. It is evident form the eqs (\ref{sq}) and (\ref{sq1})
that the Hamiltonian in eq (\ref{hal}) is decoupled into tildien and non tildien parts as mentioned in eq (\ref{ji}). Hence, we work with 
one of the Hamiltonians and similar analysis goes through for the other Hamiltonian.
By considering the Hamiltonian 
\begin{eqnarray}
H_1  &=&\omega (a^{\dagger}a+ b^{\dagger}b)+\frac{i\kappa\Delta(t)}{2}
\left(ab - a^{\dagger}b^{\dagger}\right)\nonumber\\&&-\frac{U\Delta(t)}{2}(a^{\dagger}b^{\dagger}+ab),
\label{sqy}
\end{eqnarray}
and applying the following transformation among the mode operators
\begin{eqnarray}
A&=&\mu a + \nu^* b^\dagger, \; A^\dagger = \mu^* a^\dagger + \nu b\label{bo1}\\
B&=&\mu b + \nu^* a^\dagger, \; B^\dagger = \mu^* b^\dagger + \nu a\label{bo2}
\end{eqnarray}
with $\mu=e^{i\phi_\mu}\vert \mu\vert$ and $\nu=e^{i\phi_\nu}\vert \nu\vert$, ($\phi_\mu, \phi_\nu \in R$), we diagonalize the Hamiltonian $H_1$. 
Similar analysis goes through for the other Hamiltonian $H_2$. With a bit of algebra  for $\kappa=0$,
one can exactly diagonalize the Hamiltonian (\ref{sqy}) by using the transformations (\ref{bo1}) and (\ref{bo2}) . Let  $\kappa=0$ for $t=0$,  then we evolve the state through the schr\"odinger (\ref{hal}) at this time. Then the corresponding non-tildien Hamiltonian $H_0$ is given by
 \begin{eqnarray}
H_0  &=&\omega (a^{\dagger}a+ b^{\dagger}b) -\frac{U\Delta(t)}{2}(a^{\dagger}b^{\dagger}+ab),
\label{sqh}
\end{eqnarray}
and the final diagonalized Hamiltonian $H_f$ (which is diagonalized version of $H_0$), after the unitary transformation (\ref{bo1}) and (\ref{bo2}),  is given by
\begin{eqnarray}
H_f=S^{-1}(r)H_0 S(r)=\Omega^2[A^\dagger A + B^\dagger B + 1]
\end{eqnarray}
where 
\begin{eqnarray}
S(r)=exp[r\mathcal{K}_- - r^*\mathcal{K}_+]=exp[r a^\dagger b^\dagger - r^*ab]
\end{eqnarray}
here $\Omega^2=4\omega^2-U^2\Delta^2(0)$ and $r$ is related to $\mu$ and $\nu$ in eq (\ref{bo1}) and (\ref{bo2}) 
via the following Bogolyubov coefficients :
\begin{eqnarray}
\mu=cosh(r)&=&\frac{\omega}{\sqrt{\omega^2-\frac{U^2\Delta^2(0)}{4}}},\nonumber\\
\nu=sinh(r)&=&\frac{U\Delta(0)}{2\sqrt{\omega^2-\frac{U^2\Delta^2(0)}{4}}}.
\end{eqnarray}
Note that here
\begin{eqnarray}
\vert \mu \vert^2-\vert \nu \vert^2=1.
\end{eqnarray}
This fixes $\Delta^2(0)=\frac{\omega^2+1}{U^2}=$Constant. The solution to the Schr\"odinger equation 
\begin{equation}
i\hbar\frac{\partial}{\partial t}\vert \psi_0(t)\rangle=-iH_0\vert \psi_0(t)\rangle
\end{equation}
where Hamiltonian $H_0$ is given by eq (\ref{sqh}),
is given by
\begin{eqnarray}
\vert\psi_0(t)\rangle &=& exp[r\mathcal{K}_- - r^*\mathcal{K}_+]\vert\psi_0(0)\rangle\nonumber\\
&=& exp[r a^\dagger b^\dagger - r^*ab]\vert\psi_0(0)\rangle.
\end{eqnarray}
Similar solution exists for the $\tilde{H}_0$ where
\begin{eqnarray}
\tilde{H}_0  &=&\omega (\tilde{a}^{\dagger}\tilde{a}+ \tilde{b}^{\dagger}\tilde{b}) -\frac{U\Delta(t)}{2}(\tilde{a}^{\dagger}\tilde{b}^{\dagger}+\tilde{a}\tilde{b}).
\label{sqhtil}
\end{eqnarray}
\subsection{Self Consistency Analysis}
After achiving the solution of the Schr\"odinger (\ref{hal}) without the non-linearity (as say $\kappa=0$), we now compute the background field $\Delta(t)$ self-consistently by taking the initial state of the two-mode system to be the vacuum state.
The aim is to reinforce the non-linearity. In thermo field dynamic notation the two-mode vacuum state is given by $\vert \rho (0)\rangle=\vert 00, \tilde{0}\tilde{0}\rangle$, while in the 
usual notation $\rho(t)=e^{-i\int dt\hat{H}}\vert 0,0\rangle \langle 0,0\vert e^{i\int dt\hat{H}}$. In thermo field dynamics, the 
background field $\Delta(t)$ is computed as expectation value of the averaged creation and annihilation operators, and is given by
\begin{equation}
\Delta(t)=\langle ab \rangle=\langle I\vert ab \vert \rho(t)\rangle = Tr (ab\rho(t)).
\end{equation}
After doing a bit of algebra (see Appendix B) one gets  $\Delta(t)$ to be
\begin{eqnarray}
\Delta(t) &=& (1 + i\omega t)C \Delta(0) +\left(1+\bar{n}_1\bar{n}_2\right)\frac{(U-i\kappa)}{2}\int_0^t dt' \Delta(t')sinh^2(C\Delta(0)t).\label{sco}
\end{eqnarray}
where $c=\frac{U}{\omega}$. This is nothing but the Fredholm integral equation of the second kind. Computing the integral upto first order one gets
\begin{eqnarray}
\Delta(t)&=& (1 + i\omega t)C \Delta(0)-\left(1+\bar{n}_1\bar{n}_2\right)\frac{(U-i\kappa)(1-i\omega)}{2}[\frac{\lambda t}{2}-\frac{\lambda}{2\Omega}  sinh(C\Delta(0) t)].
\end{eqnarray}
By considering  $\Omega t$ to be small (so that $sinh(C\Delta(0) t)\simeq C\Delta(0) t)$ one gets
\begin{eqnarray}
\Delta(t)= (1 + i\omega t)C \Delta(0).\label{del}
\end{eqnarray}
Hence the solution of the Schr\"odinger equation (\ref{hal}) is given by eq (\ref{soll}), which when written in terms of the
$su(1,1)$ generators, becomes
\begin{eqnarray}
\vert \rho(t) \rangle &=& \Big(exp[\zeta_{a3}\mathcal{ N} +\zeta_{a-}\mathcal{K}_{-}
+ \zeta_{a+} \mathcal{ K}_{+}]\\ \nonumber&&\otimes
exp[\zeta_{b3} \tilde{\mathcal{ N}}+\zeta_{b-} \tilde{\mathcal{K}}_{-} + \zeta_{b+}\tilde{\mathcal{K}}_{+}]
\Big)\vert \rho (0)\rangle,
\label{fi12}
\end{eqnarray}
where $\zeta_{a3}=i\omega t$, $\zeta_{a-}=\int dt \frac{\Delta(t)}{2}(iU+\kappa)$, $\zeta_{a+}=\int dt \frac{\Delta(t)}{2}(iU-\kappa)$,
$\zeta_{b3}=i\omega t$, $\zeta_{b-}=\int dt \frac{\Delta(t)}{2}(iU+\kappa)$ and $\zeta_{b+}=\int dt \frac{\Delta(t)}{2}(iU-\kappa)$.
Using the disentanglement formula \cite{pervo} one can write eq (\ref{fi12}) as
\begin{eqnarray}
\vert \rho(t)\rangle &=& \big\{\left(exp[\Gamma_{a+} \mathcal{ K}_{+}]exp[ln(\Gamma_{a3} \mathcal{ N}]
 exp[\Gamma_{a-} \mathcal{ K}_{-}]\right)\otimes\nonumber\\&&
\left(exp[\Gamma_{b +} \tilde{\mathcal{ K}}_{+}] exp[ln(\Gamma_{b 3} \tilde{\mathcal{ N}})]
exp[\Gamma_{b-} \tilde{\mathcal{ K}}_{-}]\right)\big\}\nonumber\\&& \vert \rho(0)\rangle.\label{sot1}
\end{eqnarray}
where
\begin{eqnarray}
\Gamma_{i\pm}&=&\frac{2\zeta_{i\pm}sinh\phi_i}
{2\phi_i cosh\phi_i-\zeta_{i3}sinh\phi_i},\nonumber\\
\Gamma_{i3}&=&\frac{1}
{\left( cosh\phi_i-\frac{\zeta_{i3}}{2\phi_i}sinh\phi_i\right)^2}
\end{eqnarray}
with
\begin{eqnarray}
\phi_i^2&=&\frac{\zeta_{i3}^2}{4}-\zeta_{i+}\zeta_{i-}\label{phl}
\end{eqnarray}
and $i$ stands for $a$ and $b$.

It can be clearly seen that the $\Gamma_i$'s are functions of the background field $\Delta(t)$.
By using the expression for $\Delta(t)$ from eq (\ref{del}) (which is valid for small $t$) we get
\begin{eqnarray}
\Gamma_{i\pm}&=& \frac{ \Delta(0)t}{2}(iU\pm\kappa)(1 +  \frac{\omega^2 t^2}{4}).
 \end{eqnarray} 
 By taking $(iU\pm\kappa)=-\zeta e^{i\phi}$ then one gets
 \begin{eqnarray}
\Gamma_{i\pm}&=& -\frac{ \Delta(0)\zeta t}{2}(1 +  \frac{\omega^2 t^2}{4})e^{\pm i\phi}.\label{sor1}
 \end{eqnarray} 
 
By considering any arbitrary initial state
$\rho(0)=\sum_{m,n}^{\infty}\rho_{m,m',n.n'}\vert m,m'\rangle\langle n,n' \vert$ of a single mode system
where $\vert m, m'\rangle$ and $\vert n,n'\rangle$ are number of states, in the thermo field dynamic notation, there by using eq(\ref{10}) (of appendix D), $\vert \rho(0)\rangle$ takes the form
\begin{eqnarray}
 \vert \rho(0)\rangle=\sum_{m,n}^{\infty}\rho_{m,n}(0)\vert m,m',n,n'\rangle\label{rio1}.
\end{eqnarray}
Putting eq (\ref{rio1})in (\ref{sot1}) one gets

\begin{widetext}
\begin{eqnarray}
\rho_{m,n}(t)&=&
\sum_{q'=0}^{min(m',n')}\sum_{p'=0}^\infty
\left[\left(\begin{array}{c}
m'+p'-q' \\ p'
\end{array}\right)\left(\begin{array}{c}
n'+p'-q' \\ p'
\end{array}\right)\left(\begin{array}{c}
m' \\ q'
\end{array}\right)\left(\begin{array}{c}
n' \\ q'
\end{array}\right)\right]^\frac{1}{2}\nonumber\\&&\times
\sum_{q=0}^{min(m,n)}\sum_{p=0}^\infty
\left[\left(\begin{array}{c}
m+p-q \\ p
\end{array}\right)\left(\begin{array}{c}
n+p-q \\ p
\end{array}\right)\left(\begin{array}{c}
m \\ q
\end{array}\right)\left(\begin{array}{c}
n \\ q
\end{array}\right)\right]^\frac{1}{2}\nonumber\\&&
 \nonumber\\&&\times[\Gamma_{a +}]^{p'}[\Gamma_{a 3}]^{(m'+n'-2q'+1)/2}
[\Gamma_{a-}]^{q'}[\Gamma_{b +}]^{p}[\Gamma_{b 3}]^{(m+n-2q+1)/2}
[\Gamma_{b -}]^{q}\nonumber\\&&\times
 \rho_{m+p-q,m'+p'-q',n+p-q+n'+p'-q'}(0).
 \end{eqnarray}
\end{widetext}

\section{Entanglement}
The solution of the Schr\"odinger equation (\ref{hal}) is a pure state in the thermo field dynamic notation 
and is given by
\begin{eqnarray}
\vert \rho(t)\rangle &=& \big(exp[\Gamma_{a+} \mathcal{ K}_{+}]exp[ln(\Gamma_{a3} \mathcal{ N}]
 exp[\Gamma_{a-} \mathcal{ K}_{-}]\nonumber\\&&\times
exp[\Gamma_{b +} \tilde{\mathcal{ K}}_{+}] exp[ln(\Gamma_{b 3} \tilde{\mathcal{ N}})]
exp[\Gamma_{b-} \tilde{\mathcal{ K}}_{-}]\big)\nonumber\\&& \vert \rho(0)\rangle.
\end{eqnarray}
where $\Gamma_{i\pm}$ are given in eq (\ref{sor1}) and $K_\pm$ are given in eq (\ref{k1}) and (\ref{k2}). In this Schr\"odinger equation, as there is no mixing 
between the tildian and non-tildian modes in the Liouville space, one can write the solution in the system Hilbert 
space as
\begin{widetext}
\begin{eqnarray}
\rho(t) &=& \big(exp[\Gamma_{a+} \mathcal{ K}_{+}]exp[ln(\Gamma_{a3} \mathcal{ N}]
 exp[\Gamma_{a-} \mathcal{ K}_{-}]\big) \rho(0)
\big(exp[\Gamma_{a+} \mathcal{ K}_{+}]exp[ln(\Gamma_{a3} \mathcal{ N}]
 exp[\Gamma_{a-} \mathcal{ K}_{-}]\big)\label{syu}
\end{eqnarray}
\end{widetext}
where again $\Gamma_{i\pm}$ are given in eq (\ref{sor1}) and $\rho(0)$ is the initial state of two mode system.  One can clearly see that this a two-mode squeezed state
of the two-mode Hilbert space. It is well known that two-mode squeezing gives rises to entanglement. As an example, we take the initial state $\rho(0)$ to be the two-mode thermal state. Then, to calculate the 
entanglement of the time evolved state $\rho(t)$ we go over to phase space description by following transformation
\begin{eqnarray}
a &=&\frac{1}{\sqrt{2}}(x + ip_x), \;
a^\dagger =\frac{1}{\sqrt{2}}(x - ip_x), \label{ean}\\
b &=&\frac{1}{\sqrt{2}}(y + ip_y), \;
b^\dagger =\frac{1}{\sqrt{2}}(y - ip_y). \label{ean1}
\end{eqnarray}
Putting them in the eq (\ref{syu}) one gets the two-mode squeezed thermal state in the phase space
corresponding to the two-mode the real squeezing transformation  
\begin{equation}
S(r)=\left(\begin{array}{cccc}
cosh(r) & 0 &sinh(r) &0\\
0 & cosh(r) & 0 & -sinh(r)\\
sinh(r) & 0& cosh(r) &0 \\
0 & -sinh(r) &0 &cosh(r)
\end {array}\right),
\end {equation}
where the squeezing parameter is given by 
\begin{eqnarray}
r=\frac{ \Delta(0)}{2}(1 +  \frac{\omega^2 t^2}{4})\zeta t.\label{rty}
\end{eqnarray}
 The covariance matrix of a  two-mode thermal state is given by
\begin{equation}
\sigma=n_1\mathbb{1} \oplus n_2\mathbb{1}
\end {equation}
where the $n_1$ and $n_2$ are sympletic eigenvalues of the covariance matrix. The covariance matrix for the state
$\rho(t)$ in eq (\ref{syu})is given by
\begin{equation}
V=S(r)\sigma S^\dagger(r)
\end{equation}
and so
\begin{equation}
V=\left(\begin{array}{cccc}
p & 0 &-s &0\\
0 & p & 0 & s\\
-s & 0& q &0 \\
0 & s &0 &q
\end {array}\right)
\end {equation}
with $p=n_1cosh^2(r) + n_2 sinh^2(r)$, $q= n_1 sinh^2(r)+n_2cosh^2(r) $ and
$s=\pm \frac{n_1 + n_2}{2}sinh(2r)$.

So the covariance matrix is of the form
\begin{equation}
V=\left(\begin{array}{cc}
A & C \\
C^T & B
\end {array}\right)
\end {equation}
where
\begin{eqnarray}
A=\left(\begin{array}{cc}
p & 0 \\
0 & p 
\end {array}\right),\; B=\left(\begin{array}{cc}
q & 0 \\
0 & q 
\end {array}\right),\nonumber\\C=\left(\begin{array}{cc}
-s & 0 \\
0 & s 
\end {array}\right),\; J=\left(\begin{array}{cc}
0 & 1 \\
-1 & 0 
\end {array}\right)
\end {eqnarray}

Then the separability condition \cite{sim} for any two-mode state reads as
\begin{widetext}
\begin{equation}
detAdetB +\left(\frac{1}{4}-\vert detC\vert\right)^2-tr(AJCJBJC^TJ)\geq \frac{1}{4}(detA+detB)\label{sep}
\end{equation}
\end{widetext}
So, the state $\rho(t)$  in eq (\ref{syu}) is entangled if and only if the condition  in eq (\ref{sep}) is satisfied. Then the thermal state is entangled iff 
\begin{equation}
sinh^2(r)\ge \frac{(n_2^2-1)(n_1^2-1)}{(n_1 + n_2)^2}.
\end{equation}
For $n_1=n_2=n$, the amount of entanglement in $\rho(t)$ is given in terms of logarithmic negativity
\begin{equation}
E_N(r)=-\frac{1}{2}[log_2(e^{-4r}/n)].\label{eny}
\end{equation}
\section{Decoherence}
As we have seen the previous section, that the solution to $\rho(t)$ of the master equation (\ref{pair}) in the Hilbert 
space $H$ is given by eq (\ref{syu}) and 
$\Gamma_{i\pm}$ are given in eq (\ref{sor1}). To calculate decoherence effects of $\rho(t)$
we compute $\rho^2$ and is given below. Note that
\begin{eqnarray}
\rho(t) &=& exp[-\Delta(0){2}(1 +  \frac{\omega^2 t^2}{4})\zeta t] \rho(0)\label{syp}
\end{eqnarray}
Then 
\begin{eqnarray}
Tr[\rho^2(t)]&=&Tr[\sum_{m,n}\langle m,n\vert\rho^2(t)\vert m,n\rangle ]\nonumber\\&=& exp[-\Delta(0){2}(1 +  \frac{\omega^2 t^2}{4})\zeta t] \label{syo}
\end{eqnarray}
The behaviour of decoherence  is plotted fig. 2. One can see immediately that for the short time itself as the value of damping coefficient increases the system decoheres faster.  
\begin{figure}[hb]
\begin{center}
\includegraphics[height=2in,width=2.8in]{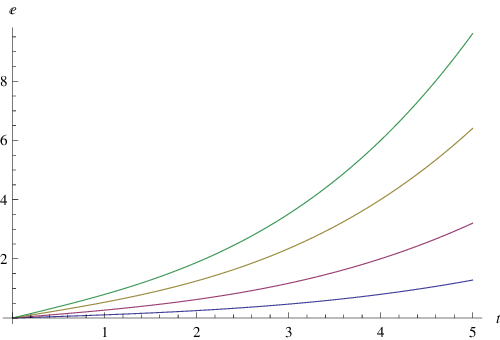}
\caption{Time vs Entanglement: Here $\omega=.25$, $\Delta(0)=.25$ and $n=.5$ the curves from bottom to top corresponds to $\zeta=  0.1,~ 0.25,~0.5~ and~ 0.75$. We use eq (\ref{rty}) and  eq (\ref{eny})}
\end{center}
\end{figure} 
\begin{figure}[hb]
\begin{center}
\includegraphics[height=2in,width=2.8in]{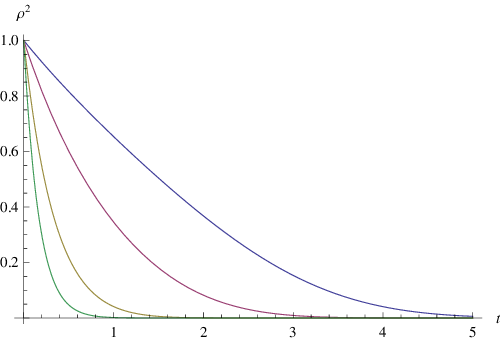}
\caption{Time vs Decoherence: Here $\omega=0.25$ $\Delta(0)=0.25$ curves from top to bottom corresponds to $\zeta=  1,~ 0.75,~0.5~ and~ 0.25$. }
\end{center}
\end{figure} 

\section{Conclusion}
In this paper, the techniques of thermo field dynamics and the 
Hartree-Fock approximation are used to solve the master equation for the special case of two-mode  Bose-Hubbard model
in the presence of a non-linear damping. We have treated the approximated field self-consistently and computed it for a small time $t$ analytically.  Then, the decoherence and the entanglement for this system are computed. We show that the entanglement for this system for a short time increases when the initial state is in two-mode thermal state. We interpret this behaviour due to the existence of the non-linear medium. To get the exact picture for the behaviour of the entanglement for a long time, one has to do the numerical studies. For the large time $t$ we expect the initial state to damp much faster due to the presence of non-linear damping. It can be seen form the decoherecnce plot that as the value of the damping coefficient increases the damping in the system is faster, as expected. We expect that the further numerical studies using this model will give better results and these results may be applied to condensed matter systems.

\section{Appendix}
\subsection{Appendix A: Calculation of Bogolyubov Coefficients} 

Consider the following Hamiltonian
 \begin{eqnarray}
H_0  &=&\omega (a^{\dagger}a+ b^{\dagger}b) -\frac{U\Delta(t)}{2}(a^{\dagger}b^{\dagger}+ab),
\label{sqhf}
\end{eqnarray}
By applying the following Bogolyubov or squeezing transformation
\begin{eqnarray}
A&=&\mu a + \nu^* b^\dagger, \; A^\dagger = \mu^* a^\dagger + \nu b\nonumber\\
B&=&\mu b + \nu^* a^\dagger, \; B^\dagger = \mu^* b^\dagger + \nu a\nonumber
\end{eqnarray}
(where $\mu=e^{i\phi_\mu}\vert \mu\vert$ and $\nu=e^{i\phi_\nu}\vert \nu\vert$), we diagonalize the Hamiltonian $H_0$. Let us first 
rewrite the Hamiltonian in a convenient form as
\begin{equation}
H_0=[ a^\dagger ~~b]\left( \begin{array}{cc}
m & n\\
n & m
\end{array}\right)\left[ \begin{array}{c}
a\\
b^\dagger
\end{array}\right]\label{h}
\end{equation} 
 where $m=i\omega$ and $n=\frac{-iU\Delta(0)}{2}$.
 By identifying the coefficient matrix appearing in the eq (\ref{h}) with
 \begin{equation}
\left( \begin{array}{cc}
m & n\\
n & m
\end{array}\right)=\left( \begin{array}{cc}
cosh(r) & sinh(r)\\
sinh(r) & cosh(r)
\end{array}\right),
\end{equation} 
and with a suitable normalization $\sqrt{\vert m\vert^2-\vert n\vert^2}$, the Bogolyubov coefficients can be read off as follows
\begin{eqnarray}
\mu=cosh(r)&=&\frac{\omega}{2\sqrt{\omega^2-\frac{U^2\Delta^2(0)}{4}}},\label{mu}\\
\nu=sinh(r)&=&\frac{U\Delta(0)}{2\sqrt{\omega^2-\frac{U^2\Delta^2(0)}{4}}}.\label{nu}
\end{eqnarray}
The above two quantities satisfy
\begin{eqnarray}
\vert \mu \vert^2-\vert \nu \vert^2=1.
\end{eqnarray}
\subsection{Appendix B: Computation of $\Delta(t)$}
Consider the following field variable appeared in eq (\ref{sq}) and (\ref{sq1})
\begin{equation}
\Delta(t)=\langle ab \rangle
\end{equation}
The initial state being the thermal state 
in usual notation 

\noindent
$\rho(t)=\sum_{n,m=0}^\infty\frac{\bar{n}_1^n}{(\bar{n}_1+1)^{n+1}}\frac{\bar{n}_2^m}{(\bar{n}_2+1)^{m+1}}e^{-i\int dtH}\vert n,m\rangle \langle n,m\vert e^{i\int dtH}$ where $H$ is is equal to $H_1$ in eq  (\ref{sq}). So we have
\begin{eqnarray}
\langle ab \rangle &=& \langle I\vert ab \vert \rho(t)\rangle = Tr (ab\rho(t))\nonumber\\
&=& \sum_{n,m=0}^\infty\frac{\bar{n}_1^n}{(\bar{n}_1+1)^{n+1}}\frac{\bar{n}_2^m}{(\bar{n}_2+1)^{m+1}}
Tr(ab e^{-i\int dtH}\vert n,m\rangle \langle n,m\vert e^{i\int dtH})\nonumber\\
&=&\sum_{n,m=0}^\infty\frac{\bar{n}_1^n}{(\bar{n}_1+1)^{n+1}}\frac{\bar{n}_2^m}{(\bar{n}_2+1)^{m+1}}\sum_{p,q}\langle p,q \vert ab e^{-i\int dtH}\vert n,m\rangle \langle n,m\vert e^{i\int dtH} \vert p,q \rangle \nonumber\\
&=&\sum_{n,m=0}^\infty\frac{\bar{n}_1^n}{(\bar{n}_1+1)^{n+1}}\frac{\bar{n}_2^m}{(\bar{n}_2+1)^{m+1}}\sum_{p,q}\langle n,m \vert  e^{i\int dtH}\vert p,q\rangle \langle p,q\vert abe^{-i\int dtH} \vert n,m \rangle\nonumber\\
&=&\sum_{n,m=0}^\infty\frac{\bar{n}_1^n}{(\bar{n}_1+1)^{n+1}}\frac{\bar{n}_2^m}{(\bar{n}_2+1)^{m+1}}\langle n,m \vert e^{i\int dtH}(\sum_{p,q}\vert p,q\rangle \langle p,q\vert) ab e^{-i\int dtH} \vert n,m \rangle\nonumber\\
&=&\langle 0,0 \vert e^{i\int dtH} ab e^{-i\int dtH} \vert 0,0 \rangle +\sum_{n,m=1}^\infty\frac{\bar{n}_1^n}{(\bar{n}_1+1)^{n+1}}\frac{\bar{n}_2^m}{(\bar{n}_2+1)^{m+1}}\langle n,m \vert e^{i\int dtH} ab e^{-i\int dtH} \vert n,m \rangle\nonumber
\end{eqnarray}
Consider
\begin{eqnarray}
 e^{i\int dtH} ab e^{-i\int dtH}&=&(1+i\int dtH) ab (1-i\int dtH) =(ab+i\int dt[H, ab] )
\end{eqnarray}
Then one has
\begin{eqnarray}
\langle ab \rangle &=&\langle 0,0 \vert ab\vert 0,0\rangle + i\int dt \langle 0,0 \vert[H, ab]  \vert 0,0 \rangle\nonumber\\&&+\sum_{n,m=1}^\infty\frac{\bar{n}_1^n}{(\bar{n}_1+1)^{n+1}}\frac{\bar{n}_2^m}{(\bar{n}_2+1)^{m+1}}\left(\langle n,m \vert ab \vert n,m \rangle +i\int dt \langle n,m \vert[H, ab]  \vert n,m \rangle\right)\nonumber\\
\end{eqnarray}
The first term comes from the time independent Hamiltonian then one has 
from the eq (\ref{mu}) and (\ref{nu}) one gets 
\begin{eqnarray}
\langle ab \rangle &=&C \Delta(0) + i\int dt \langle 0,0 \vert[\zeta_{a3}\mathcal{ N} +\zeta_{a-}\mathcal{K}_{-}
+ \zeta_{a+} \mathcal{ K}_{+}, \mathcal{K}_{-}]  \vert 0,0 \rangle\nonumber\\&&
+i\sum_{n,m=1}^\infty\frac{\bar{n}_1^n}{(\bar{n}_1+1)^{n+1}}\frac{\bar{n}_2^m}{(\bar{n}_2+1)^{m+1}} \int dt \langle n,m \vert[\zeta_{a3}\mathcal{ N} +\zeta_{a-}\mathcal{K}_{-}
+ \zeta_{a+} \mathcal{ K}_{+}, \mathcal{K}_{-}]  \vert n,m \rangle
\end{eqnarray}
where $C=\frac{U}{\omega}$, we neglect $\Delta^2(0)$ term.
The above equation is by rearranging eq (\ref{sq11}). By using the commutation relations of $su (1,1)$ in eq (\ref{comu}) and
then applying the inverse squeezing transforamtions of eq (\ref{bo1}) and (\ref{bo2})
one gets
\begin{eqnarray}
\langle ab \rangle &=&C \Delta(0) + i\int dt \omega \mu\nu +\int dt \zeta_{a+}\vert \mu\vert^2 +
\sum_{n,m=1}^\infty\frac{n\bar{n}_1^n}{(\bar{n}_1+1)^{n+1}}\frac{m\bar{n}_2^m}{(\bar{n}_2+1)^{m+1}}\int dt \zeta_{a+}\vert \mu\vert^2
\\
\Delta(t)=\langle ab \rangle &=&C \Delta(0) + i\int dt \omega sinh(r)cosh(r)  
\nonumber\\&&+\left(1+\sum_{n,m=1}^\infty\frac{n\bar{n}_1^n}{(\bar{n}_1+1)^{n+1}}\frac{m\bar{n}_2^m}{(\bar{n}_2+1)^{m+1}}\right)\frac{(iU-\kappa)}{2}\int dt \Delta(t)sinh^2(r)\\
\Delta(t)=\langle ab \rangle&=&C \Delta(0) + i\int dt \omega sinh(r)cosh(r)  
+\left(1+\bar{n}_1\bar{n}_2\right)\frac{(iU-\kappa)}{2}\int dt \Delta(t)sinh^2(r)\label{pli}
\end{eqnarray}
The second and third term are from the time dependent Hamiltonian then the squeezing parameter $r$ in eq (\ref{mu}) and (\ref{nu}) becomes time dependent and $\Delta(0)$ also becomes time dependent $\Delta(t)$. We neglect $\Delta^2(t)$ term. 
Then substituting the value of $\Delta(t)$ in eq (\ref{pli}) in the squeezing parameter $r$ and keeping upto the first term  one gets
\begin{eqnarray}
\Delta(t)=\langle ab \rangle &=&(1 + i\omega t)C \Delta(0) +\left(1+\bar{n}_1\bar{n}_2\right)\frac{(iU-\kappa)}{2}\int dt \Delta(t)sinh^2(r).
\end{eqnarray}
 Then one has
\begin{eqnarray}
\langle ab \rangle &=&(1 + i\omega t) C\Delta(0)  +\left(1+\bar{n}_1\bar{n}_2\right)\frac{(iU-\kappa)}{2}\int dt \Delta(t)sinh^2(C\Delta(0)t).
\end{eqnarray}

\subsection{Appendix C: Calculation of $\phi$ and $\Gamma$}
One can see that $\Gamma_i$'s are given in terms of $\zeta_{a3}=i\omega t$, $\zeta_{a-}=\int dt \frac{\Delta(t)}{2}(iU+\kappa)$, $\zeta_{a+}=\int dt \frac{\Delta(t)}{2}(iU-\kappa)$,
$\zeta_{b3}=i\omega t$, $\zeta_{b-}=\int dt \frac{\Delta(t)}{2}(iU+\kappa)$ and $\zeta_{b+}=\int dt \frac{\Delta(t)}{2}(iU-\kappa)$.
Considering only for small time $t$, i.e., neglecting $\Delta^2(t)$ in eq (\ref{phl}) and higher order terms one has  
\begin{equation}
\phi_i^2=\frac{\zeta_{i3}^2}{4}=-\frac{\omega^2 t^2}{4},
\end{equation}
 which gives
 \begin{eqnarray}
\Gamma_{i\pm}&=&\frac{2\zeta_{i\pm}sinh(\frac{i\omega t}{2})}{i\omega t( cosh(\frac{i\omega t}{2})+ sinh(\frac{i\omega t}{2}))}\nonumber\\&=&
\frac{2\zeta_{i\pm}sin(\frac{\omega t}{2})}{\omega t}e^{-\frac{i\omega t}{2}}.\label{mk}
 \end{eqnarray}
 
Here we have used $cosh(x)+ sinh(x)=e^x$ and $-isinh(ix)=sin(x)$.
Again considering $\omega t$ to be small, that is using $sin(\omega t)=\omega t$, one has from eq(\ref{mk}):
 \begin{eqnarray}
\Gamma_{i\pm}&=&=\zeta_{i\pm} (1-\frac{i\omega t}{2})
 \end{eqnarray} 
Thus 
  \begin{eqnarray}
\Gamma_{i\pm}&=&\int dt \frac{\Delta(t)}{2}(iU\pm\kappa)(1-\frac{i\omega t}{2}).\label{mk1}
 \end{eqnarray} 
Then putting the value of $\Delta(t)$ from eq(\ref{del}) into eq(\ref{mk1}) one gets
 \begin{eqnarray}
\Gamma_{i\pm}&=&\int_0^t dt' \frac{\Delta(t')}{2}(iU\pm\kappa)(1-\frac{i\omega t}{2})\nonumber\\&=&\int_0^t dt' \frac{(1 + i\omega t') \Delta(0)}{2}(iU\pm\kappa)(1-\frac{i\omega t}{2})\\
&=& \frac{ \Delta(0)}{2}(iU\pm\kappa)(t + i\omega \frac{t^2}{2})(1-\frac{i\omega t}{2})\nonumber\\&=&\frac{ \Delta(0)t}{2}(iU\pm\kappa)(1 +  \frac{\omega^2 t^2}{4}).
 \end{eqnarray} 

 \subsection{Appendix D: Thermo Field Dynamics}

A brief description of TFD is given below. 
The dissipative term in any master equations  makes it difficult
to apply the usual Schr\"odinger equation techniques (with pure states) to mixed states.
The thermo field dynamics (TFD) provides such a formalism. 
In TFD, the mixed state averages are expressed as scalar products and the dynamics is given in terms of Schr\"odinger like equation. A density operator $\rho=\vert N\rangle\langle N\vert$ corresponding to a Fock state $\vert N\rangle$ in the Hilbert space ${\cal H}$ is viewed in TFD as a vector $\rho=\vert N, \tilde{N} \rangle$ in an extended Hilbert space ${\cal H}\otimes {\cal H}^*$. The central idea  in TFD is to construct a density operator $\vert \rho^\alpha\rangle,  1/2  \le \alpha \le 1$ as a  vector in the extended Hilbert space ${\cal H}\otimes {\cal H}^*$. 

Here the averages of operators with respect to $\rho$ reduces to a scalar product:
\begin{eqnarray}
\langle A\rangle = Tr [A\rho] &=& \langle\rho^{1-\alpha}\vert A\vert \rho^\alpha\rangle,
\end{eqnarray}
where $\vert \rho^\alpha\rangle$ is given by
\begin{equation}
\vert \rho^\alpha\rangle = \hat{\rho}^\alpha\vert I\rangle\;, \texttt{with},\; \hat{\rho}^\alpha = \rho^\alpha \otimes I,
\end{equation}
where $\vert I\rangle$ is the resolution of the
identity
\begin{equation}
\vert I\rangle=\sum \vert n\rangle\langle n\vert = \sum \vert n\rangle\otimes\vert \tilde{n}\rangle \equiv \sum \vert n,\tilde{n}\rangle,
\end{equation}
in terms of a complete orthonormal basis  $\{\vert n\rangle\}_{n=0}^\infty$  in  ${\cal  H}$.
The state vector $\vert I\rangle$ takes a normalized vector to another
normalized vector in the extended Hilbert space ${\cal H}\otimes {\cal 
H}^*$. The matrix $A(a,a^\dagger)$  acts like $A \otimes I$. 

It may be noted that for any density operator the states $\vert \rho^\alpha\rangle,  1/2  \le \alpha \le 1$ have a finite norm in the extended Hilbert space ${\cal H}\otimes {\cal H}^*$. This is not in general true for the state $\vert \rho^{1-\alpha}\rangle,  1/2  \le \alpha \le 1$, which includes $\vert I\rangle$. These state are regarded as formal but extremely useful constructs.

The creation and the annihilation 
operators $a^\dagger, \tilde{a}^\dagger,  a$,  and  $\tilde{a}$ are introduced as 
follows
\begin{eqnarray}
a\vert n,\tilde{m}\rangle &=& \sqrt{n} \vert n-1,\tilde{m}\rangle,\nonumber\\ 
a^\dagger\vert n,\tilde{m}\rangle &=& \sqrt{n+1} \vert n+1,\tilde{m}\rangle,\\
\tilde{a}\vert n,\tilde{m}\rangle &=&  
\sqrt{m} \vert n,\tilde{m-1}\rangle,\nonumber\\ \tilde{a}^\dagger\vert n,\tilde{m}\rangle 
 &=& \sqrt{m+1} \vert n,\tilde{m+1}\rangle.
\end{eqnarray}
The operators $a$ and $a^\dagger$  commute  with  $\tilde{a}$  and 
$\tilde{a}^\dagger$.  It is clear from the above that
$a$ and $a^\dagger$ acts on the vector space $\cal{H}$ and  $\tilde{a}$  and 
$\tilde{a}^\dagger$ acts on vector space $\cal{H^*}$. From the 
expression for $\vert I\rangle$ in terms of the number states
\begin{equation}
\vert I\rangle = \sum_n \vert n,\tilde{n}\rangle,
\end{equation}
it follows that
\begin{equation}
a\vert I\rangle=\tilde{a}^\dagger \vert I\rangle,\; a^\dagger\vert I\rangle = 
\tilde{a}\vert I\rangle,\label{10}
\end{equation}
and hence for any operator $A$ (written in terms of $a$ $a^\dagger$ and their complex conjugates), one has
\begin{equation}
A\vert I\rangle = \tilde{A}^\dagger \vert I\rangle,\label{11}
\end{equation}
where $\tilde{A}$ is obtained from $A$ by making the  replacements using the 
tilde  conjugation  rules   $a\to   \tilde{a},   a^\dagger   \to 
\tilde{a}^\dagger,  \alpha\to  \alpha^*$. 
An immediate consequence of this is that  the state $\vert \rho^\alpha\rangle$
remains unchanged under the replacements of $a \to  \tilde{a}$,  
$a^\dagger  \to \tilde{a}^\dagger,$  and c number $\to$ complex conjugates C by applying the  the hermiticity property of $\rho$
i.e. $\rho^\dagger=\rho$. 
The tildian property reflects the hermiticity property of the density operator.

The evolution of a conservative system in terms of $\rho^\alpha$ is given by the von Neumann equation
\begin{eqnarray}
 \frac{\partial}{\partial t}\rho^\alpha(t)=\frac{-i}{\hbar}[H,\rho^\alpha],
\end{eqnarray}
and by applying $\vert I\rangle$ from the right, one gets
\begin{eqnarray}
\frac{\partial}{\partial t}\vert \rho^\alpha(t)\rangle = -i\hat{H}\vert \rho^\alpha\rangle,\label{scr}
\end{eqnarray}
where
\begin{eqnarray}
 -i\hat{H}=i(H-\tilde{H}).
\end{eqnarray}
In TFD, one can derive a Schr\"odinger like equation for any state $\vert\rho^\alpha\rangle$
with arbitrary value of $\alpha$.   
For dissipative systems, the evolution equation is given by master equation 
\begin{eqnarray}
 \frac{\partial}{\partial t}\rho(t)=\frac{-i}{\hbar}(H\rho-\rho H)+L\rho,\label{mast}
\end{eqnarray}
where $L$ is the Liouville term. The non-equilibrium thermo filed dynamics is developed and 
analysed in $\alpha=1$ representation. Hence, from now on, we work in, $\alpha=1$ representation \cite{vs3, vs4, kap}. In this representation, for any hermitian operator $A$, one has 
\begin{equation}
\langle A\rangle = \langle I\vert A\vert \rho\rangle = \langle A\vert\rho\rangle= Tr(A\rho).
\end{equation} 
By applying $\vert I\rangle$ to the eq (\ref{mast}) from the right one goes over to TFD and the corresponding Schr\"odinger  equation is given by (\ref{sc}), 
with $-i\hat{H}$ being a tildian, and thus the problem of solving master equation is reduced to solving a Schr\"odinger like equation namely eq (\ref{sc}). 

Historically the thermo field dynamics was developed in $\alpha=\frac{1}{2}$ representation. In this representation  $ \vert \rho_0^\frac{1}{2}\rangle$ is related to the $\vert 0,0\rangle$ by a unitary transformation,
which is nothing but the Caves-Schumaker state, for details ref \cite{U1,U2,U3,U4,U5,kap,sim1}.

\end{document}